# Flexibility and regularity of the hydration structure of a sodium ion. Nonempirical insight

Yulia V. Novakovskaya

Moscow State University, Leninskie Gory 3/1, Moscow, 119991 Russia

E-mail: *jvn@phys.chem.msu.ru*

**Abstract:** The stationary nonempirical simulations of $Na^+(H_2O)_n$ clusters with n in a range of 28 to 51 carried out at the density functional level with a hybrid B3LYP functional and the Born-Oppenheimer molecular dynamics modeling of the size selected clusters reveal the interrelated structural and energetic peculiarities of sodium hydration structures. Surface, bulk, and transient configurations of the clusters are distinguished with the different location of the sodium nucleus (close to either the spatial center of the structure or one of its side faces) and its consistently changing coordination number (which typically equals five or six). The $<r_{NaO}>$ mean Na-O distances for the first-shell water molecules are found to depend both on the spatial character of the structure and the local coordination of sodium. The $<r_{NaO}>$ values are compared to different experimental estimates, and the virtual discrepancy of the latter is explained based on the results of the cluster simulations. Different coordination neighborhoods of sodium are predicted depending on its local fraction in the actual specimens.

## INTRODUCTION

Hydration of ions, especially alkaline metal ions which regulate numerous intra-, inter-, and extracellular processes in living organisms and are present in various current sources and electrolyte baths of different purposes, was repeatedly studied for a long time, and it may seem now that everything is already clear. However, there are still many questions open, primarily because of the visible closeness of some averaged characteristics, such as hydration numbers or radii of the first coordination shells of the ions retrieved from different experimental data at different assumptions or estimated in various theoretical simulations with the use of distinctly different approaches and models. The numerical differences themselves, as well as their existence, are very important, because the local coordination of ions affect their ability (the apparent energy change in the process and its activation barrier) to be coordinated to an electrode surface (or localized within the electric double layer close to the electrode surface) before discharge or to be bound to the mouth of a particular membrane protein channel followed by the transfer inside the channel, and so on. It is obvious that in all the situations, one or another kind of the reorganization in the hydration shell of the ion should take place, and its efficiency (the actual rate and yield) depends on the initial state of the ion surrounded with the solvating molecules. For example, according to a hydration mimicry concept, the ease of the ion transfer via membrane protein channels is provided by the similarity in the spatial and charge characteristics of the coordinating channel segments formed by proper combinations of amino acid residues to those typical of the ion surrounded with water molecules (see [1] and references therein).

In this respect, sodium ion is a very interesting and concurrently important example, because the ion plays a very substantial role in and even determines numerous biochemical processes, and at the same time it is known not to affect the water structure as strongly as do the so-called typical cosmotropic and chaotropic ions, such as its close neighbors in the Periodic table, namely, lithium and potassium respectively. An additional reason for the study of sodium ion is an apparent or virtual discrepancy of the old and recent experimental estimates concerning the effective radius of the first solvation shell of the ion. As quoted below, recent investigations suggest a more compact shell. Is the result caused by the improved spatial and temporal resolution of the techniques or is there any intrinsic factor that may predetermine the divergence of the experimental estimates?

What do we know about the first hydration shell of a sodium ion? There are different methods of the experimental investigation of the ionic hydration structures; and different methods especially in the early days, but (as recent papers show) in our days also, provide different results. Static structures are retrieved from the diffraction studies, where scattering of neutrons or X-rays by the specimens of interest is considered. The approaches enable one to determine the interatomic (internuclear) distances; and here neutron scattering (NS) supplements the X-ray diffraction (XRD) as providing information about positions of water protons and, hence, the orientation of water molecules around ions. Spectroscopic techniques, such as X-ray absorption fine structure (EXAFS) are applied for estimating the short-range intermolecular interactions, since the absorbance of X-rays depends on the close neighborhood of the atoms. The latter methods also supplement XRD as applicable to the solutions of a relatively low concentration, in which the XRD signals of solutes are almost lost because of the overwhelming signals of the prevailing solvent molecules. At the same time, EXAFS probes mainly the local structure and cannot provide sufficiently full information e.g. about the second hydration shell.

Furthermore, the results of any experimental measurement are processed and interpreted in one or another way; and much depends on the scheme selected for processing the raw data. Particularly, interpreting the diffraction data can be based on different models. For example, it could be assumed that the close neighborhood of an ion is symmetric, while the arrangement of more distant water molecules is either asymmetric typical of a bulk water structure (*sa* model) or also symmetric to a certain extent (*ss* model); or that there is no particular symmetry in both the hydration structure and the arrangement of surrounding water molecules (*aa* model). As was summed up in [2] for NaCl solutions of very close concentrations (characterized by the water-to-salt molar ratios of 10.2 and 13.9), the interpretation of XRD data on the basis of *sa* and *ss* models provided the coordination number (*CN*) of sodium of 6 and 4 respectively at comparable estimates of the mean Na-O distances of 2.41 and 2.42 Å. The same *CN* values (namely, 4 and 6) were obtained in terms of *aa* model for the solutions of another nature, namely NaI and $NaNO_3$ of also close but lower water-to-salt molar ratios of 7 and 6.1-9.3 respectively. How can and should one interpret the results? A priori it seems that the model where no special assumption about the structure symmetry is invoked should be more adequate, but the difference seems too large. At the same time, typical mean coordination numbers determined from XRD studies of moderately concentrated aqueous solutions of sodium salts were 4.9 in 4m $NaClO_4$ [3], and 5.1 in 3m NaCl [4]. Moreover, there is seemingly no preferable number of water molecules in the next shells: high resolution time-of-flight mass spectrometry of the aquacomplexes of $Na^+$ ions produced by electrospray ionization in gas phase [5] revealed no "magic numbers" of water molecules in the clusters, which involved up to a total of 80 molecules. Both kinds of the latter results support the idea about the absence of pronounced symmetry in the arrangement of water molecules at short- and long-range distances from sodium. However, the question about the prevailing number of the first-shell molecules, their separation from the ion and orientation in space remains if not completely unsolved, but still open. For example, quite recent XRD experiments [6] provided the *CN* value of sodium of 5.5±0.3 and 5.9±0.6 in 6m and 2.6 m NaCl solutions in a reasonable agreement with the previous investigations, but smaller mean Na-O distances of 2.384±0.003 Å (and even smaller, 2.370±0.024 Å based on the EXAFS data) compared to the aforementioned ones. It is worth noting that the *CN* value of sodium ion obtained as a result of the EXAFS data processing was characterized by a large uncertainty, 5.4±1.3. This result may reflect the actual coexistence of hydration structures with different numbers of molecules in the first solvation shell of a sodium ion though the scattering from 4 to 7 at first glance seems too large. Here, it is necessary to mention a very recent paper [7], in which the dynamic Stokes radius of a sodium ion was estimated to be 2.44 Å in aqueous NaCl solutions independent of the concentration in a range of 0.5 to 6 M based on the data of the pulsed field gradient NMR method. Note that the value characterizes the hydration of sodium on a time scale about 1 μs, which is much larger compared to typical XRD or EXAFS scales; and the result obtained reflects the time-averaged size of the first hydration shell of sodium.

If we now turn to the theoretical predictions, there are three kinds of simulations in this field, namely, (i) molecular dynamics (MD) studies of molecular ensembles, (ii) Monte Carlo (MC) simulations; and (iii) quantum chemical cluster modeling. Actually it was modeling of relatively small $Na^+(H_2O)_n$ clusters that provided parameters for constructing potentials at the early stage of MD and MC studies. At first, calculations of water dimer and ion-water pair at the Hartree-Fock level supplemented with corrections for the electron correlation [8, 9] were used for parameterizing a Bernal-Fowler-like model potential, which was then applied to MC modeling of $Na^+(H_2O)_n$ clusters with n up to 6. Later on, different but generally similarly constructed potentials were used already for periodic simulations of dilute aqueous solutions [10, 11] and provided *CN* estimates in a broad range of 4.3 (and a radius of 2.59 Å) to 5.96±0.02 respectively. In the end of 1980s, Clementi group has shown [12] that such analytical water-water and water-ion potentials result in the overrated sodium ion hydration energy by about 21%, though the overall trends observed for a series of ions were consistent with the experiments and the Born theory.

By that time, it was already shown and realized that polarizable water models [13] and many-body potential functions [14, 15] can provide a better description of small $Na^+(H_2O)_n$ clusters. For example, a discovered 5+1 configuration of $Na^+(H_2O)_6$ cluster [15] was later generalized as a *CN* value of sodium between 5 and 6 in the MC simulations of an NPT ion-water ensemble [16]. The efficiency of polarizable nonadditive many-body potentials, which include Lennard-Jones and electrostatic interactions in water-water and water-ion pairs, as well as a nonadditive polarization energy and a three-body exchange repulsion, in modeling both the gas-phase $Na^+(H_2O)_n$ clusters and the ion solvation in liquid water was confirmed also in MD simulations [17] where the mean *CN* value of sodium was found to be 5.7. A close range of 5 to 6 (with no clear difference between the intramolecular geometries and molecular dipole moments of the molecules in the first and second solvation shells) was predicted in the MC simulations with refined ab initio-based potentials [18]. Later on, nonempirical simulations of $Na^+(H_2O)_n$ clusters with n up to six carried out at the Moeller-Plesset second order perturbation theory (MP2) level with double and triple-zeta basis sets [19] showed that nonadditive interactions are actually large, and reliable results can be obtained when up to four-body terms are taken into account in the model potentials. A close conclusion followed from the combined quantum-classical (QM/MM) simulations [20] where the first solvation shell of the ions was treated at the Hartree-Fock (HF) level, while the residual water molecules, with the use of a specially parameterized potential: the determined mean *CN* value of 5.6±0.3 was lower compared to that obtained with pair potentials (6.5±0.2). Furthermore, the effective fragment potential (EFP) approach was also found to give results for the sodium hydration in an agreement with the HF simulations [21]. When the quantum chemical level was substantially improved (MP2 in a resolution identity variant and DFT with the PBE generalized gradient functional and a basis set of split valence polarized triple-zeta quality) and used for the description of the twelve inner-shell molecules [22], the sodium *CN* was found to be again in nearly the same range of 5.7-5.8. Finally, when nonempirical molecular dynamics in Car-Parrinello formulation became available, it (with the PBE functional and a double zeta basis set) predicted a very close estimate of the mean number of water molecules in the first solvation shell of sodium, namely 5.2 [23]. At the same time, many averaged structural and energetic characteristics of the aqueous ionic systems, even such as ion mobilities [24], ionic association in aqueous solutions [25], or general properties of solutions of different concentrations and even liquid solution-vapor interfaces [26–29] were found to be reproduced with a sufficient accuracy (compared to experiments) even with unpolarizable potentials. For example, MC approach used in combination with a rigid (TIP4P) water potential and a combination of Lennard-Jones and Coulomb contributions to the ion-water interaction potential [30] predicted the mean *CN* value of sodium of 5.8 at the approximate insensitivity of the first hydration shell to the concentration of a model NaCl solution. However, this nearly proper mean first-shell configuration was observed alongside with the overestimated enthalpy of the water-shell formation around the ion, which (as mentioned above) is the known general drawback of the pair-wise additive potentials (like TIP4P) compared to those, which explicitly include many-body effects (like SPCE/POL) [31, 32]. Concurrently, effective three-body and polarizable potentials were discovered to predict similar first-shell structures; and only the shortening of the mean distance between the first and second solvation shells was noticed when the polarizable potentials were used [33].

This is a good illustration of the quite natural idea that potentials fitted to a set of experimentally measured properties or their combination with nonempirically estimated values for the ion-water interactions can provide close and quite reliable mean characteristics of the complex aqua systems, such as radial distribution functions or hydration free energies of the ion [34, 35]. But what can be said about the local parameters of the systems? Averaged as well as integral values may be the same, while the differential characteristics may differ. Furthermore, as the above brief overview of the works in the field shows, an apparently correct coordination number can be found independent of the concentration of a model ion-containing system (which seems scarcely possible at the proper description of the hydration structures inevitably sensitive to the neighborhood) and/or accompanied with the overrated hydration energies or enthalpies (which reflects certain drawbacks in reproducing contributions of different nature to the total interaction energy of the system). In fact, these peculiarities mean that in empirical models, some intrinsic errors in the energy contributions can be leveled off by each other at certain configurations to provide reasonable minimum-energy structures, but become substantial at some others to give finally incorrect difference values. In the case of purely nonempirical studies, if one restricts the scope of the reference systems to some relatively small number of hydrating molecules (e.g., a dozen or even typically six), effects that are predetermined by the H-bond network formed by water molecules (in fact, collective rather than simply many-body effects) can be lost, and, hence, cannot be reproduced when the model constructed is applied to the description of large systems. Therefore, nonempirical simulations of relatively large $Na^+(H_2O)_n$ clusters with n up to half a hundred seem a sole way to clarifying the physical aspects of the ion hydration at a micro (actually, nano) level.

**METHODICAL**

To clarify the possible reasons for the differences in the experimental estimates of the local structural parameters of the hydration shells of sodium, paying special attention to the discrepancy between the previous and more recent [6] data, we have undertaken nonempirical investigation of $Na^+(H_2O)_n$ clusters with n up to 55. As noted above, it seemed necessary to extend information about the configuration of sodium aquacomplexes beyond small typical n=6 (which corresponds to the first solvation shell solely) [36, 37] and medium-sized clusters with n up to 16 or 20 (which correspond to nearly one and a half shells around the ion) [38, 39]. The necessity can be founded as follows. When one considers a small $Na^+(H_2O)_n$ cluster with n up to 6, all the molecules are nearly uniformly arranged around the central ion, and in a majority of stable configurations there is no hydrogen bond between them. However, it is H-bonds that predetermine most of the characteristics (either structural or energetic) of water-based cluster systems. Undoubtedly, additional water molecules, even if they do not noticeably change the energetically preferable spatial orientation of those molecules, which form the first solvation shell, affect the ability of the latter to share their electron density (or in other words, to be involved in the localization of the positive charge of the system). Furthermore, water molecules, similarly to other typical H-bond forming particles, are known to demonstrate collective properties when bound to each other, and the most pronounced collective effects are observed when the molecules are involved in closed H-bond sequences (rings), within which covalent and hydrogen bonds are alternating, e.g. (H)O-H...(H)O-H...(H)O-H... We suggested to call such sequences conjugated, because such alternation provides spatial possibilities for the actual conjugation of the bonds due to the substantial $\pi$-kind contribution to the typical H-bonding [40]. Examining the orientation of water molecules around a sodium ion, it can be noticed that due to the predominant ion-dipole interactions, both O-H bonds of each water molecule are oriented from the ion (at a certain tilt angle) and form H-bonds with additional water molecules, so that the former ones can scarcely be involved in the aforementioned closed conjugated H-bond sequences. However, the molecules in the second hydration shell already can be involved in such rings and in this way provide the indirect correlation between the states (e.g., vibrational) of the first-shell molecules. The larger the number of the molecules beyond the first coordination shell, the higher the mutual correlation between the molecules and the larger the collective effects.

Another related aspect that deserves particular attention is the relative number of molecules at different sides of the ion, or in other words, variations in the position of the ion, which should not necessarily be a central particle of the structure. An asymmetry in the coordination shell may take place in different situations. These include (i) the dynamic reorganization of any actual solution (even with a relatively low or moderate concentration) when ions can approach each other from time to time to form close solvent separated pairs; (ii) the migration of ions under the effect of an external field when they finally come close to an adsorbing electrode surface or an absorbing mouth of a membrane protein channel where their partial dehydration should inevitably take place; and (iii) the reversible thermal transfer of ions toward the solution surface. While the second listed situation takes place only under special conditions at a directed ion transfer governed by the external electric field gradient, the two other variants are typical of any actual solution, including those in experimental samples studied with the use of XRD or EXAFS techniques. Then, a question arises whether the existence of ions with asymmetric hydration shells can noticeably affect the measured averaged values, such as the coordination numbers or radial distribution functions; and if yes, then how.

Here, we would like to draw attention again to the fact that information retrieved from any experiment reflects the averaged characteristics of specimens, which is natural for we deal with dynamically changing subjects; and even a picosecond range covers continuous repeated changes in the positions of nearly all the nuclei of the system. This can be treated as an averaging or the corresponding broadening of the signal due to the thermal motions. Therefore, we face a virtual dilemma. To reproduce features of a liquid-phase specimen, one should use dynamic modeling of its supposedly constituting fragments of smaller or larger size, but the modeling should be based on the nonempirical (adiabatic) potentials rather than those parameterized to reproduce experimentally measured mean values. Reliable conclusions can be based only on the sufficiently broad scope of data for a variety of clusters, but at this final point, averaging again masks individual intrinsic characteristics of clusters and can give the same final result for different sets of initial data. At the same time, it is the individual characteristics that can be obtained from the stationary simulations of clusters, but the latter are believed to be only idealized precursors of the real structures. Below, we are going to show that stationary simulations of water-based clusters provide the kernel data of the thermally varying objects and enable one to judge the structural characteristics of vibrationally averaged fragments and, hence, explain experimental data.

Model clusters considered in this work were constructed in two ways. In one series of structures, these were generated by a gradual increase in the number of molecules at their nearly uniform arrangement around the ion. Such structures are referred to as *bulk* clusters below. In the other series, the structures were generated by removing a part of water molecules from an optimal bulk cluster configuration at one side of the ion and either completely excluding them from the system or adding to the residual sides of the ion solvation shell in an arbitrary manner. Such structures are referred to as *surface*. In both variants the initially generated structures were then optimized and a normal-coordinate analysis was used to confirm that the resulting configuration corresponds to an energy minimum of the adiabatic potential. Of the clusters of a particular composition of each type, only one with the lower energy and the most extended H-bond network at the reasonable compactness (see below) was selected for the further analysis. Additionally, *transient* clusters were considered at selected compositions, which either resembled bulk ones in a general character of the structure (primarily, sodium position close to the spatial center of the system), but had the same $CN$ number of sodium as the surface ones or, by contrast, resembled the surface configuration with the higher $CN$ number.

Conventional statistical thermodynamic calculations were carried out to estimate the vibrational contribution to the enthalpy and Gibbs energy of the cluster under normal conditions (at a room temperature and a pressure of 1 atm). Knowing these thermal energy contributions (denoted as $H_{vib}$ and $G_{vib}$ below) makes it possible to compare the overall stability of the structures of the same composition but different arrangement of water molecules. Additionally, relative enthalpies and Gibbs energies ($H_{rel}$ and $G_{rel}$) of the clusters of the same composition were estimated by taking into account the difference in their total electronic energy, the lowest energy being set as a reference (zero) level. Furthermore, insofar as the corresponding relative enthalpy and Gibbs energy characterize the thermal stability of the H-bond network of the cluster, $H_{rel}$ and $G_{rel}$ values normalized to the number of water molecules ($H_{rel}/n$ and $G_{rel}/n$) were also estimated for all the configurations considered that makes it possible to compare clusters of different molecular sizes.

Comparing the individual clusters of a relatively large molecular size to those immersed in the aqueous environment or to the instantaneous fragments of the dynamically changing environment, it can be noticed that the difference between the corresponding vibrational contributions is mainly restricted to the surface molecules, whose oscillations (either stretching of OH bonds uninvolved in hydrogen bonds or swinging) proceed with larger amplitudes and other frequencies. And at the same number of water molecules, the difference in the estimated thermal contributions (related to the vibrational motions only) reflect (i) the degree of ordering and the collectivity of oscillations of the internal water molecules of the cluster and (ii) the difference in the number of surface water molecules with incomplete neighborhoods. The role of the latter factor can be estimated by comparing clusters with the known difference in the number of such undercoordinated molecules.

In the analysis of the resulting bulk and surface configurations of $Na^+(H_2O)_n$ clusters, a special attention was paid to the local coordination of sodium and the mean Na-O distances estimated for the molecules involved in the first coordination shell. Additionally, the structural peculiarities of the whole clusters were considered, particularly the numbers of molecules with different coordination neighborhoods, which not only reflect the character of the H-bond network of the cluster, but also can be considered as a sign of the reasonableness of the structures obtained, because the undercoordinated molecules can be involved in H-bonding with the surrounding layers of molecules in an actual specimen. By undercoordinated we imply those molecules, which form less than four hydrogen bonds, two as donors (*d*) and two as acceptors (*a*) of H-bond protons. Accordingly, a so-to-say ultimately perfect coordination neighborhood is of *ddaa* kind; but naturally in any actual dynamically evolving system, there should be quite a large number of tri-coordinate molecules, and there may appear penta-coordinate ones as well. Therefore, all the systems were characterized by a scope of numbers of molecules of the predominant coordination types (*N*(*ddaa*), *N*(*dda*), *N*(*daa*)). Here, we tried to find a correlation between the structural kinds of clusters, the coordination number of sodium (*CN*), and the mean Na-O distances in the first hydration shell.

To prove the validity of the analysis based on the information about the optimum thermally unexcited configurations of individual clusters, a series of dynamic simulations were also carried out for a selected cluster. The simulations were based on the Born-Oppenheimer approximation (BOMD); and the electronic problem was solved at the same level as the one used for the stationary calculations (see below). Taking into account that most of the experiments are carried out at a room or close temperature, the evolution of a cluster was initiated by the excitation of those normal vibrational modes that are characterized by frequencies no higher than 210 cm$^{-1}$. These are typically either quite local or relatively delocalized (covering large structure segments) oscillations that distort the mutual arrangement of molecules, particularly change the distances between their centers of masses and the sodium nuclei. Thus, these are the motions that mimic thermal distortions in the actual specimens. The dynamic trajectories were constructed at a time step of 0.5 fs and a total duration of 5 ps.

As integral spatial characteristics of Na$^+$(H$_2$O)$_n$ systems, $g(r_{NaO})$ radial distribution functions, which are given in most of the theoretical dynamic simulations and in many experimental studies, were also considered. These were estimated as follows:

$$g(r_{NaO}) = \frac{1}{4\pi r^2 V} \frac{\delta n_O(r)}{\delta r}$$

where $\delta n_O(r)$ is the number of oxygen nuclei in a spherical layer of $\delta r$ thickness at a distance of $r$ to $r+\delta r$ from the sodium nucleus. Here, the function is normalized to the $V$ volume of the cluster. The functions discussed below are constructed based on the data for the $r$ range of 2.2 to 5.0 Å, the latter value being equal to the mean apparent radius of the smallest clusters, so that the molecules most distant from sodium in the larger clusters (which formally belong to rarefied layers) were excluded from the analysis. The $\delta r$ step was selected as 0.05 Å, which is quite reasonable to provide a sufficiently correct estimate of the layer volume as $4\pi r^2 \delta r$ for the distances larger than 2.2 Å. The $V$ volume and $S$ surface values were found as the volume and surface area of the overlapping van-der-Waals spheres of all the atoms involved in the system with r(H) = 1.2 Å, r(O) = 1.5 Å, and r(Na) = 2.1 Å.

The method of nonempirical simulations was selected based on the following requirements and restrictions. It should be sufficiently accurate in reproducing peculiarities of the electron density distribution in H-bonded systems and at the same time reliably efficient to provide the possibility of modeling clusters that involve up to 55 water molecules. The second-order Moeller-Plesset perturbation theory and density functional approach with hybrid exchange-correlation functionals are known as meeting such requirements and used in the aforementioned papers [36-39]. Taking into account the more rapid increase in the calculation costs of the MP2 approach compared to the DFT even with a hybrid functional and the known better approximation of the curvature of the adiabatic potential in the vicinity of minima, we selected DFT with a B3LYP functional. An additional argument in favor of the choice was the necessity to carry out BOMD simulations, for which a variational approach (such as DFT) is preferable compared to a perturbation one for the possibility of using Hellmann-Feynman forces solely with no need for the correction by adding Pulay forces. In what concerns the basis set, on one hand, it should be sufficiently flexible (comparable to those used for the description of relatively small clusters) and, on the other hand, it should not be linearly dependent (as a usual requirement to those used in periodic simulations). As numerous aforementioned simulations showed, basis sets of double zeta quality are sufficient for describing clusters of interest, but it is necessary to include polarization functions in the basis set always and to extend the set with diffuse functions especially when modeling relatively small clusters. Insofar as the objects of our investigation are clusters that involve no less than 28 molecules, the extra diffuse functions can only deteriorate the results because of the partial linear dependence of the basis functions. In the central part of the cluster, which is of the primary importance to us, there is a sufficient number of tails of the functions centered on the neighboring and even relatively distant nuclei to provide a correct description of the electron density distribution. Therefore, 6-31G(d,p) basis set was consistently selected for all the atoms. All the calculations were carried out with the use of Firefly 8.2 program package [41] and visualized with the Chemcraft [42] software.

**RESULTS AND DISCUSSION**

From the very beginning it should be stressed that the coordination number of sodium (five or six) is not a sign or consequence of a particular character of the cluster, namely surface or bulk. Among the systems considered, there are bulk clusters with $CN$=5 and surface clusters with $CN$=6. Moreover, in some situations those with a penta-coordinate sodium have a lower energy, while in certain situations it is hexa-coordinate sodium that is preferable at the same total number of molecules. Below, we analyze various coordination neighborhoods of sodium and derive certain conclusions about the possible prevailing kinds of sodium hydration under different physical conditions.

Both surface and bulk configurations of $Na^+(H_2O)_n$ clusters with $n$ = 28, 33, 38, 43, 46, and 51 were considered. Judging from the $g(r_{NaO})$ radial distribution functions (see e.g. [1, 43, 44]), the first peak Na-O distances are no larger than 3.0 Å (here the $g(r_{Na-O})$ function has the minimum), while the second-peak distances fall in a range of 3.0 to ca. 5.7 Å. Hence, a cluster that pretends to be a model of nearly two solvation shells around sodium should be characterized by the largest Na-O distances about 5.7 Å. In our simulations, the smallest such cluster was $Na^+(H_2O)_{28}$; therefore, we do not involve smaller clusters in the discussion below. It is worth noting that the analysis of the abundance spectra of $Na^+(H_2O)_n$ clusters produced by electrospraying 0.015M aqueous NaOH solution [43] revealed the highest peak intensity for the clusters with

n=28, which undoubtedly reflects their stability probably determined by a reasonable number of water molecules in a relatively closed structure. All larger clusters correspond to an increase in the number of water molecules successively in three to five in such a way that the number of molecules with dangling OH bonds was minimal possible. Surface clusters of the same total composition, but with the initial position of sodium close to a surface of the cluster, which means that on one side of sodium, there is nearly one hydration shell, while on the residual sides, the apparent thickness of the water layer is larger, are analyzed alongside with the bulk ones. For illustration in the largest considered $Na^+(H_2O)_{51}$ cluster pair, the change in the apparent one-side thickness of the hydration structure is about 1.45 Å, which readily shows that there is a pronounced asymmetry of the hydration structure.

As a test cluster we selected $Na^+(H_2O)_{33}$ and carried out a thorough investigation of its possible configurations and their variations depending both on the initial approximation (in stationary simulations) and time (in dynamic runs). The choice was predetermined by the fact that this was the smallest cluster where a bulk-type structure with the *CN* number of 6 was found. In the smaller clusters, sodium was penta-coordinate (see below). An additional argument was the energetic closeness of the configurations of different types. The results were then used in the analysis of the larger clusters and, hence, in drawing conclusions concerning the probable coordination neighborhoods of sodium.

**Stationary vs. dynamic features of the size-selected cluster system**

$Na^+(H_2O)_{33}$ cluster turned out as the smallest one (among those studied in this work), whose discovered surface, bulk, and transient configurations differ in the number of water molecules in the first solvation shell of sodium, namely, five and six (Fig. 1), and the adiabatic potential values of the corresponding minima fall in a range of ca. 4.5 kcal/mol (Table 1), which is smaller than a typical water-water hydrogen bond energy. This fact makes the cluster a promising object for dynamic studies, since mutual transformations of the local configurations of water molecules, including those in the first solvation shell of sodium, may be possible within not as prolonged dynamic runs. The lowest-energy structure (Fig. 1a) of the cluster is the one of the bulk type with *CN*=5. The next structure in the energy scale (Fig. 1b) can be referred to as transient, because despite the larger *CN*=6 the number of molecules at one side of sodium is smaller compared to the other. The highest of the three energy values characterizes the surface configuration (Fig. 1c) with *CN*=5.

**Table 1.** The structural (the coordination number of sodium (*CN*), the mean Na-O distance for the *CN* molecules ($<r_{NaO}>$, Å); the number of hydrogen bonds ($N_{H-bonds}$); the numbers of water molecules with different coordination neighborhoods (*N(dda)*, *N(ddaa)*, and *N(daa)*); the mean number of bonds formed by one water molecule ($<N_{bonds}>$)) and energetic characteristics (the relative total electronic energies ($E_{rel}$), the vibrational contributions to the enthalpies and Gibbs energies under normal conditions ($H_{vib}$ and $G_{vib}$), the relative enthalpies and Gibbs energies ($H_{rel}$ and $G_{rel}$), and their normalized values ($H_{rel}/n$ and $G_{rel}/n$), kcal/mol) of Na$^+$(H$_2$O)$_{33}$ cluster with different configurations (see text for details)

| conf. | CN | $<r_{NaO}>$ | $N_{H-bonds}$ | N(dda) | N(ddaa) | N(daa) | $<N_{bonds}>$ |
|---|---|---|---|---|---|---|---|
| bulk | 5 | 2.369 | 53 | 12 | 8 | 12 | 3.4 |
| transient | 6 | 2.466 | 53 | 12 | 8 | 10 | 3.4 |
| surface | 5 | 2.375 | 55 | 10 | 12 | 7 | 3.5 |
| 1800-fs transient | 5 | 2.359 | 53 | 11 | 9 | 11 | 3.4 |
| 1400-fs surface | 5 | 2.365 | 55 | 10 | 12 | 10 | 3.4 |

| | $E_{rel}$ | $H_{vib}$ | $H_{rel}$ | $H_{rel}/n$ | $G_{vib}$ | $G_{rel}$ | $G_{rel}/n$ |
|---|---|---|---|---|---|---|---|
| bulk | 0 | 582.7 | 582.7 | 17.7 | 491.2 | 491.2 | 14.9 |
| transient | 3.8 | 583.2 | 587.0 | 17.8 | 490.8 | 494.6 | 15.0 |
| surface | 4.7 | 584.4 | 589.0 | 17.8 | 491.2 | 495.9 | 15.0 |
| 1800-fs transient | -2.3 | 583.6 | 581.3 | 17.6 | 493.1 | 490.8 | 14.9 |
| 1400-fs surface | -0.8 | 584.6 | 583.8 | 17.7 | 493.8 | 493.0 | 14.9 |

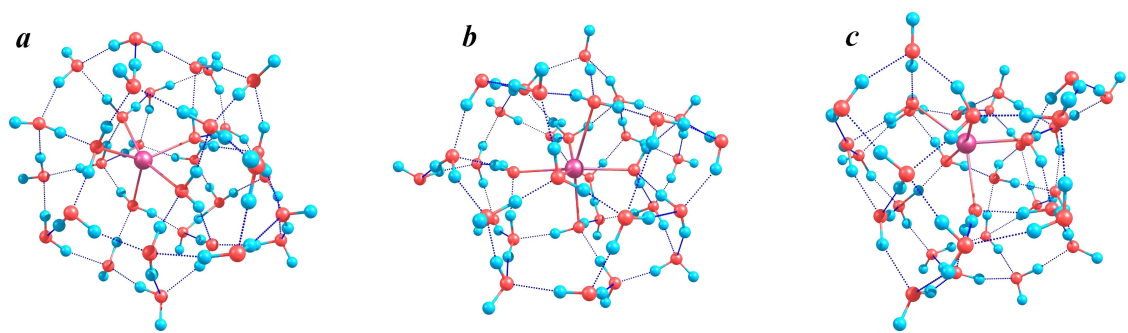

*Figure 1.* Configurations of Na$^+$(H$_2$O)$_{33}$ cluster: (a) bulk; (b) transient; and (c) surface. Coordination Na-O bonds between the sodium nucleus and the first-shell water molecules are shown with solid sticks for clarity.

Although the configurations of clusters differ substantially (see Fig. 1), their formal structure parameters, such as the total number of hydrogen bonds and the numbers of molecules with different neighborhoods are quite close. Here, it is necessary to take into account that the first-shell molecules, which are typically of *dda* coordination type in the H-bond network, are also coordinated to sodium, which makes them actually tetra-coordinate. As a result, the mean number of bonds formed by each molecule ($<N_{bonds}>$) is nearly the same for the different configurations of the cluster. Furthermore, insofar as it is the bonding character that predetermines the stability of water-based clusters, not only the adiabatic potentials, but also the thermal contributions to the energy are close (Table 1). At the same time, the bulk cluster has the lowest both enthalpy and Gibbs energy, which reflects its energetic favorableness under normal conditions. However, the differences are not as large and when normalized to the number of constituting water molecules (whose motions predetermine the thermal distortions of any actual specimen) are about 0.1 kcal/mol only. Here, the normalized relative enthalpy per molecule is about 17.7 kcal/mol. It seems reasonable to supply an excitation energy of a comparable value to the cluster to promote its dynamic changes by activating the low frequency modes.

It was the structurally transient and energetically intermediate configuration shown in Fig. 1b that was selected as the reference one for the investigation of the dynamic evolution. Additionally, dynamic trajectories were generated for the surface structure as well. The latter was predetermined by the highest energy of the initial configuration and the existence of a slightly larger number of normal modes with low frequencies.

Upon supplying a total kinetic energy of 17.2 kcal/mol to all 58 normal modes of the transient configuration with the frequencies in a range of 27.7 to 204.4 cm$^{-1}$, the following features were observed. First of all, the initial cluster becomes less compact, and sodium acquires a possibility to shift from its center (some selected snapshots are shown in Fig. 2). As a result, its

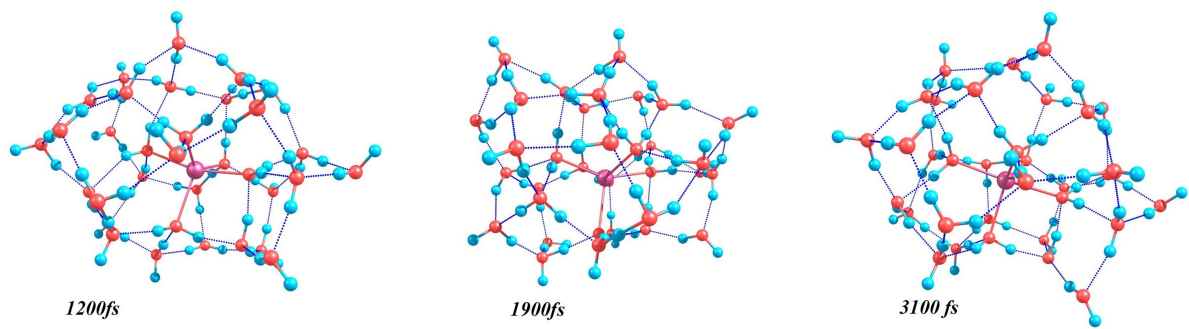

*1200fs*      *1900fs*      *3100 fs*

***Figure 2.*** Configurations of Na$^+$(H$_2$O)$_{33}$ cluster that appear in 1200, 1900, and 3100 fs of the dynamic propagation of the initially transient configuration. Coordination Na-O bonds between the sodium nucleus and the first-shell water molecules are shown with solid sticks for clarity.

neighborhood (or coordination shell) becomes less symmetric, and from about 300 fs on, the number of the molecules in its first solvation shell decreases by one, which makes the whole structure more similar to either the surface or bulk configurations from time to time (cf. Figs. 1 and 2). A Na-O distance histogram (Fig. 3a) illustrates the general trend. The histogram is based on the data of arbitrarily selected 33 configurations appeared in three 5-ps propagations. For comparison, the distances typical of the initial minimum-energy structure are also shown. Despite the fact that some changes, which take place during the dynamic run, are possible only in the individual cluster, which has no next neighbors from the outside, they are informative, since enable us to judge what kind of hydration structures can exist at this particular number of water molecules. If one looks more closely at the histogram that characterizes the changes in the inner solvation shell of sodium (Fig. 3b), it becomes clear that smaller Na-O distances become more usual in the dynamically changing system compared to the stationary optimum clusters. The selected time dependence of the mean Na-O distance in the first hydration shell (Fig. 3c), which is quite typical of numerous runs, shows that the structure is breathing (undoubtedly not uniformly, namely, when some molecules are approaching the sodium nucleus, some others are becoming more distant), and a conditional mean of the oscillations, 2.386 Å, is noticeably smaller than the initial value of the stationary configuration (2.466 Å). The amplitude of oscillations is up to 0.07 Å at the changes in particular Na-O distances as large as 0.4 Å. The mean period of oscillations (definitely anharmonic) may be estimated as ca. 300 fs. And in about 1900 fs the system acquires the possibility of drastic reorganization. The extremely high peak in $<r_{NaO}>(t)$ dependence reflects the fact that one of the five molecules of the first solvation shell drifted away from the sodium nucleus to about 3.2 Å while the residual four molecules resided at 2.358–2.472 Å. This made it possible for the molecules to rearrange around sodium; and the process was actually promoted by the preceding general increase in the apparent volume of the cluster (see Fig. 3d) which substantially increased in the previous conditional period of oscillations. This observation illustrates quite a natural delay in the inner reorganization of the cluster initiated by the appropriate loosening of the whole structure. However, it is worth noting that because of the general asymmetry of the structure, the apparent volume of the cluster changes not as largely (Fig. 3d), though (by contrast to the first solvation shell) it is always larger than the stationary value (569.2 Å$^3$). Its oscillations are shown in two variants that correspond to different virtual sensitivity of measurements (in 0.1 and 0.2 ps). As can be seen, the rougher estimates only loose some intermediate oscillations but reproduce the general trend and even the mean value (572.2 vs. 572.3 Å$^3$). The latter value is larger than those of all the stationary configurations (569.3, 569.2, and 570.5 Å$^3$ of the bulk, transient and surface structures), but the smallest volume attained during the dynamic run was at the same level (569.4 Å$^3$), while the largest (577.2 Å$^3$) exceeded it by 1% which seems quite possible under actual conditions when the fragment is surrounded with water layers. This is further supported by the changes in the apparent *S* surface area of the cluster (Fig. 3e). The oscillations in *S* value nearly follow those of *V*, which means that there is no drastic change in the shape of the system. Moreover, as can be seen, the changes in the first solvation shell of sodium are much more pronounced and cause the general contraction of the mean shell parameters.

To clarify how far can the dynamically appeared configurations be from those discovered in stationary geometry optimization runs, some structures were selected for further geometry optimization. The selection was based on the analysis of the mean Na-O distances in the first shell, general distortion of the cluster shape, and its apparent volume. Figure 4 shows two examples of this kind. One structure was the one originated upon 1800 fs of the dynamic evolution of the transient cluster at the above conditions. The other was the one originated upon 1400 fs of the evolution of the surface cluster upon supplying 19.1 kcal/mol to its 61 normal modes with frequencies of 14.7 to 206.9 cm$^{-1}$. What do we see? The resulting optimized structures (Fig. 4) are of either transient or nearly surface type with a penta-coordinate sodium nucleus located more closely to one cluster face and more distant from the others due to the generally more prolate shape of the whose structure. Characteristics of the clusters are given in Table 1. The mean Na-O distances are close or only slightly larger than the smallest one typical of the bulk configuration. The numbers of H-bonds and the numbers of molecules with different coordination neighborhoods fall in the same range as those in the clusters found in purely stationary simulations. Note that the structure referred to as 1800-fs transient one in Table 1 was structurally and energetically very close to another one, which was obtained starting from the 1500-fs configuration; and these are the two time moments that actually precede and succeed the most strongly distorted configuration in the dynamic run. The adiabatic energies are lower than that of the bulk configuration, which means that additional distortion of the structure can always provide possibilities for further improvement in the mutual arrangement of molecules. This seems quite natural for the clusters composed of such a large number of particles. At the same time, the differences are not as large, smaller than a half or even a sixth of the mean H-bond energy. However, it is worth noting that the thermal increments in the enthalpy and Gibbs energy due to the vibrational excitation of the clusters are higher and nearly or completely levels off their energetic superiority. This means that thorough search for the possible configurations of Na$^+$(H$_2$O)$_n$ clusters in stationary runs (when those most compact, with the largest mean numbers of coordination bonds of water molecules and the lowest thermal increments to the energy are selected) provides a reliable spectrum of structures that enables one to draw quite founded conclusions.

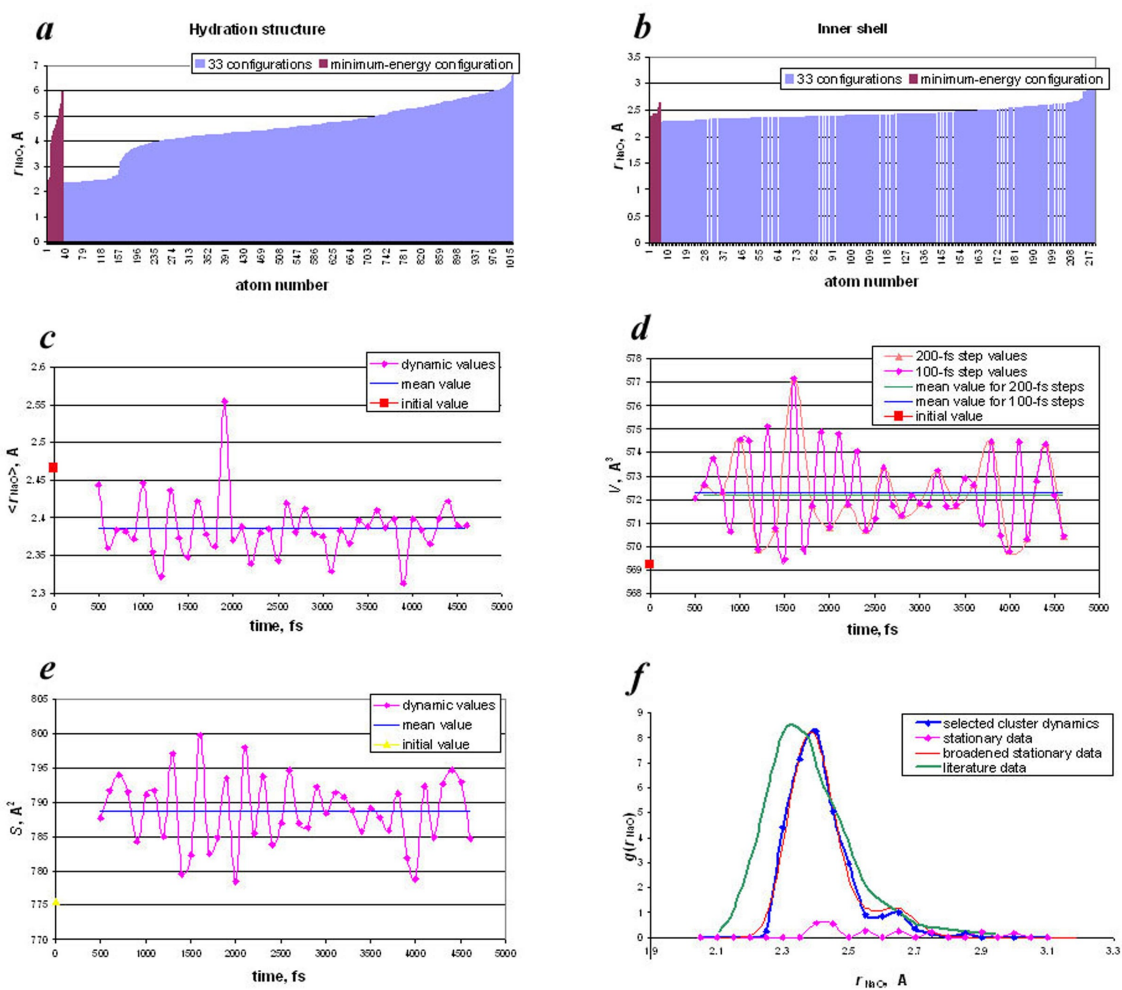

***Figure 3.*** Dynamics of the changes in the structure parameters of Na$^+$(H$_2$O)$_{33}$ cluster: (a, b) Na-O distance histograms for (a) all and (b) inner-shell water molecules constructed for 33 arbitrarily selected configurations appeared in three 5-ps runs; (c, d, e) time dependences of (c) <$r_{NaO}$>mean Na-O distance for the first coordination shell of sodium, (d) volume, and (e) surface area of the cluster that characterize its 5-ps dynamic evolution; and (f) $g(r_{NaO})$ radial distribution function constructed for tree 5-ps trajectories compared to the initial configuration of the cluster and the literature data based on the classical molecular dynamics with an empirical potential [44].

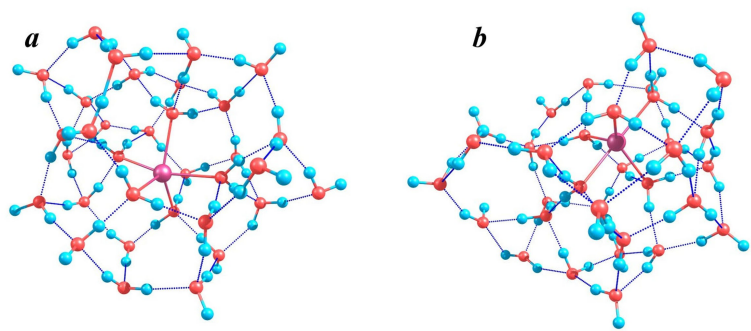

***Figure 4.*** Minimum-energy structures of Na$^+$(H$_2$O)$_{33}$ cluster obtained from the particle arrangements appeared in (a) 1800 fs of the dynamic evolution of the initially transient configurations and (b) 1400 fs of the dynamic evolution of the initially surface configuration. Coordination Na-O bonds between the sodium nucleus and the first-shell water molecules are shown with solid sticks for clarity.

In this respect, it is interesting to consider the $g(r_{NaO})$ radial distribution function constructed for the 5-ps dynamic propagations of the $Na^+(H_2O)_{33}$ transient configuration (Fig. 3f). Additionally the same plot represents a conditional $g(r_{NaO})$ function for the initial minimum-energy bulk cluster. As can be seen, the function that is based on the dynamic data can be thought of as formally broadened function of the stationary system. The broadening can be reproduced by simply centering Gaussian functions at the points where the point-wise $g(r_{NaO})$ of the individual cluster is nonzero and adding one more Gaussian at a point $r = 2.35$ Å, which reflects a substantial weight of configurations that appear due to the repeatedly approaching inner-shell molecules to sodium (Fig. 3c). This fact points how an actual distribution function can be judged based on the stationary data. At the same time, if one compares the resulting broadened function to the function that was obtained as a result of the molecular dynamic modeling of an ensemble composed of 512 water molecules and an ion with the use of a specially developed water-ion interaction model (taken from [44]), it immediately becomes clear that one cluster, whose dynamic evolution passes through various configurations, can by no means give adequate representation or description of the large-size systems. Even renormalized to the same maximum value, the functions strongly differ in a general shape and in the weight of the more compact first-shell configurations. This is what can be retrieved from the analysis of a broader set of clusters.

**Stationary features of all the clusters studied**

The general trend in the arrangement of water molecules with respect to sodium (Figs. 5 and 6) can be guessed from the Na-O distance histograms (Fig. 7a,b). First of all, there are four molecules in the surface clusters and five in the bulk ones that are always most close to sodium. The next (fifth and sixth respectively) molecule can be either more or less distant and belong formally to either the first or the second coordination shell as discussed in detail below. However, already these histograms show that the first shell in the case of surface structures is more compact, i.e., characterized by smaller Na-O distances. Furthermore, there are typically two or three molecules in a number of surface clusters that are most distant from sodium and located at Na-O distances that increase in step compared to the previous ones, which is a sign of the structure extended in one direction that can be viewed as a vertically prolate bowl if sodium is conditionally placed in the top part of the structure. Concurrently, some other surface structures are characterized by the noticeably smaller largest Na-O distances, which reflects the horizontally prolate bowl structure. Thus, on the whole, diverse variants of the structure asymmetry are reproduced in the clusters considered in this work. As to the bulk clusters, judging from the distance histograms, one can notice that there are also two kinds of them. In one of them, several molecules are more distant from sodium even compared to those in a larger cluster (cf. $Na^+(H_2O)_n$ with n= 43 and 46), which reflects that some additional molecules added to a smaller cluster can find proper positions, while some others find themselves as a protuberance, which serves as a nucleus of the subsequent coordination shell, but the number of molecules is yet insufficient for the rational formation of the shell segment. In the latter case, adding more molecules at the same side of the cluster provides the necessary minimum amount that promotes the formation of a closed structure element properly bound to the underlying shell. Again, including all the clusters in the analysis enables us to cover various structural building blocks that may affect the central part of the cluster, i.e., the local coordination neighborhood of sodium.

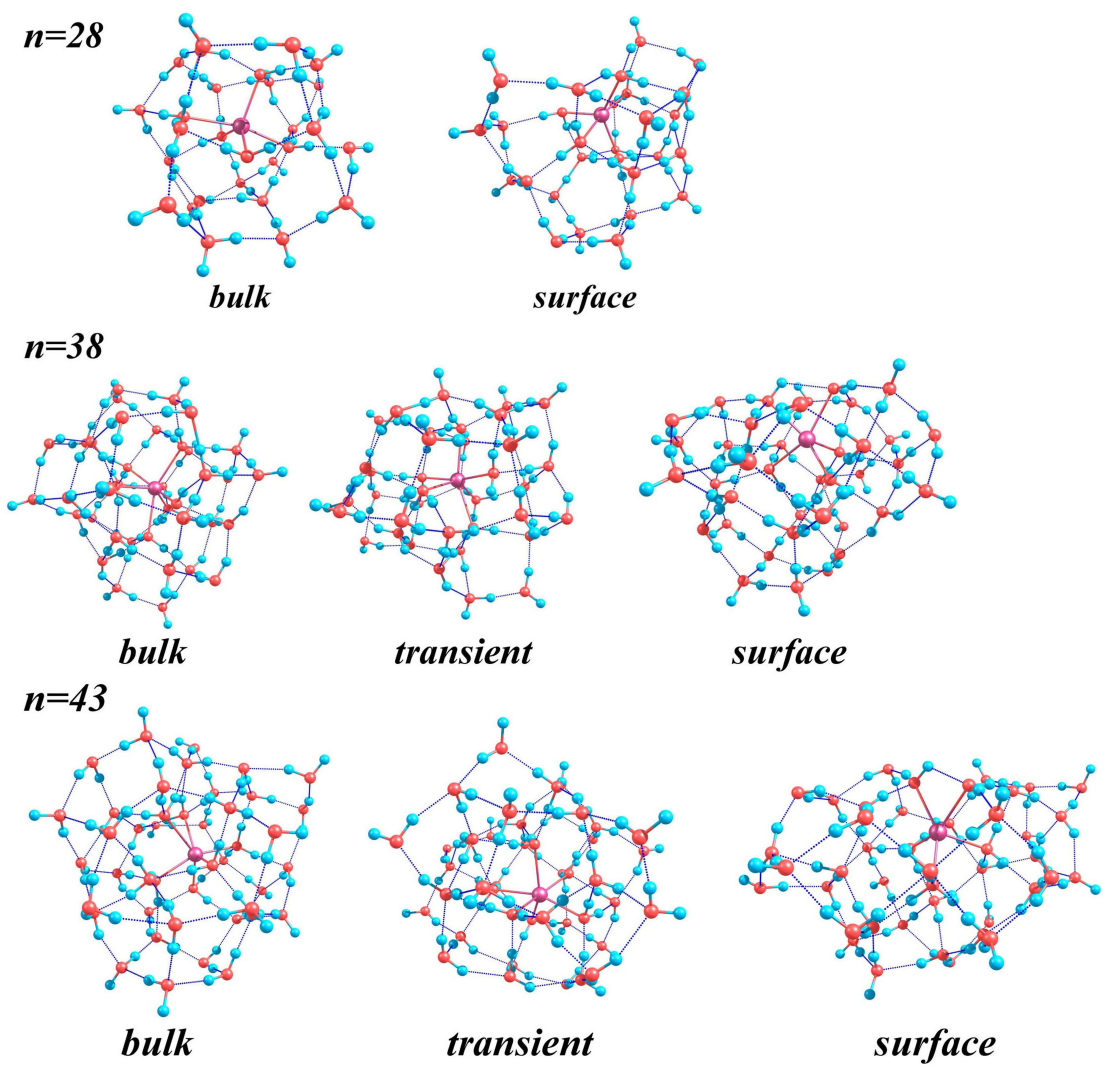

*Figure 5.* Bulk, transient, and surface configurations of $Na^+(H_2O)_n$ clusters with n=28, 38, and 43. Coordination Na-O bonds between the sodium nucleus and the first-shell water molecules are shown with solid sticks for clarity.

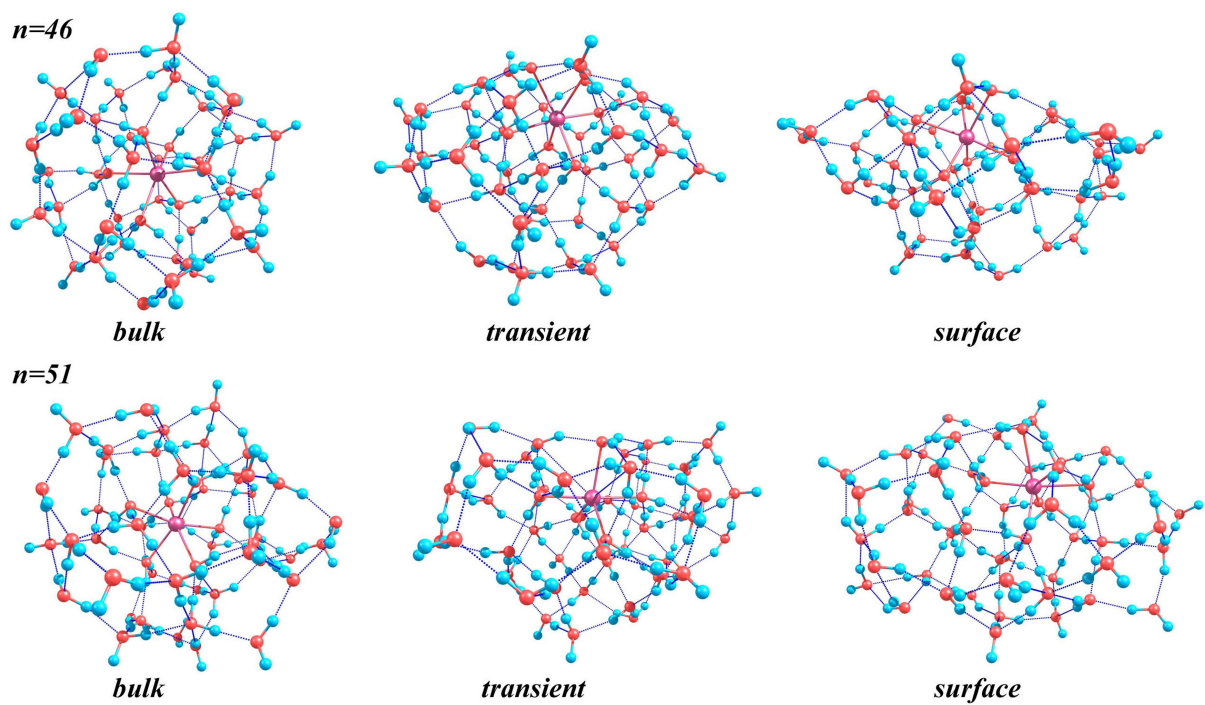

*Figure 6.* Bulk, transient, and surface configurations of $Na^+(H_2O)_n$ clusters with n=46 and 51. Coordination Na-O bonds between the sodium nucleus and the first-shell water molecules are shown with solid sticks for clarity; extra long bonds are shown with dotted lines.

At this moment, it is interesting to compare the obtained integral Na-O distance histogram for the dynamically changing $Na^+(H_2O)_{33}$ cluster (Fig. 3a), which illustrates possible arrangements of water molecules around sodium, to the histograms shown in Fig. 7 (a,b), which illustrate similar features of the minimum-energy bulk and surface $Na^+(H_2O)_n$ clusters with n = 28 to 51. In fact, integral histograms for the dynamically varying $Na^+(H_2O)_{33}$ cluster and for the whole set of bulk clusters are quite close in the general shape. However, the absolute values that characterize the first solvation shell of the dynamically varying structure are smaller (as was already mentioned above).

Therefore, it is interesting to consider the $g(r_{NaO})$ radial distribution function now. First of all, let us compare the data for the clusters of different types. Figure 7 (c,d) shows the $g(r_{NaO})$ distribution functions for the surface, bulk, and all clusters considered. As can be seen, the first-shell band of $g(r_{NaO})$ function for the bulk configurations is nearly symmetric with quite a pronounced peak similar to a cusp. In the case of the surface configurations, the band is asymmetric with a steeper left slope and a peak at a smaller $r_{NaO}$ value, which is essential in view of the above discussion (Fig. 7c). As a result, when data for the clusters of all kinds are collected in one function (Fig. 7d), one can see already a broader band with a smooth peak extended toward smaller $r_{NaO}$ values. Comparing it to the function retrieved from the experimental data [45], one can notice that now the stationary-structure $g(r)$ function has a maximum at nearly the same values as the experimental one, but the first-shell peak is narrower, though differs from the experimental ones quite symmetrically, which is natural for the absence of thermal broadening well illustrated by the above dynamic simulations of Na$^+$(H$_2$O)$_{33}$ cluster. This can be interpreted as an indication that the scope of structures involved in the analysis is already sufficient as covering most of the main coordination types of sodium in water clusters.

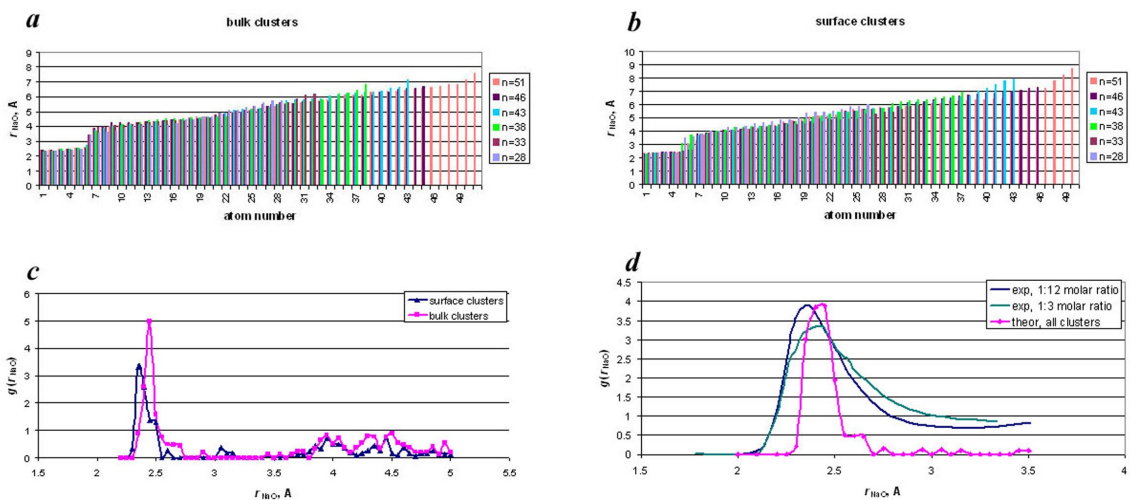

***Figure 7.*** The structure parameters of Na$^+$(H$_2$O)$_n$ clusters: (a, b) Na-O distance histograms for (a) all bulk (b) all surface configurations; (c, d) $g(r$NaO$)$ radial distribution functions constructed for (c) all bulk and all surface configurations separately and (d) all cluster configurations compared to the experimental estimates obtained for concentrated NaOH solutions with sodium-to-water molar ratios of 1:12 and 1:3 [45].

Then, we can turn to the details of the cluster systems. Coordination numbers of sodium and averaged distances between sodium and the closest oxygen nuclei, as well as the numbers of hydrogen bonds between water molecules and the numbers of molecules with different coordination neighborhoods are summed up in Table 2. At first glance there seems no regularity in the figures, but let us look at them more thoroughly. The numbers of hydrogen bonds illustrate the degree of involvement of water molecules in a continuous H-bond network of the cluster. As was noted above, each first-shell molecule forms an additional coordination bond with sodium (which is equivalent to their bonds with the proton-donating neighbors), which increases the total number of tetra-coordinate molecules. The larger the number of H-bonds, the more extended and branched the H-bond network of the cluster and, hence, the stronger the correlation in the vibrational states of the molecules, which is reflected in the $H_{vib}$ values listed in Table 3. These values should be considered along with purely geometrical parameters of the clusters and necessarily taken into account even in rough estimates of the relative energies of the clusters especially when the latter are considered as models or precursors of vibrationally averaged H-bonded liquid-phase fragments. Similarly to the information about the size selected cluster, additionally to the relative energies of the adiabatic potential values of the cluster ($E_{rel}$), their sums with $H_{vib}$ increments are given in Table 3 as $H_{rel}$ values. To make the comparison of the clusters of different molecular size more rational, the normalized $H_{rel}/n$ values are also shown to give an idea about the relative thermal energies of different hydration structures depending on the coordination of sodium and the aggregation of water molecules around it.

What can be said about the clusters considered in this work based on the data of Tables 1–3? In the case of $Na^+(H_2O)_{28}$ cluster, both surface and bulk configurations were found to be characterized by *CN*=5 (Fig. 5). At such a relatively small number of water molecules, the bulk and surface configurations do differ. The surface one is so-to-say more open, so that a sodium nucleus is only slightly countersunk. As a result, the whole configuration is strongly asymmetric with respect to sodium whose coordination is close to a trigonal bipyramid, and water molecules form a typical network characterized by five- and four-membered rings. In the case of the bulk configuration, the coordination of sodium better resembles a tetragonal pyramid, and along with four- and five-membered rings there are six- and tri-membered ones in the H-bond network as well. The latter peculiarity is predetermined by the insufficient local number of molecules when they are nearly uniformly distributed in space around sodium. As a result, the total number of H-bonds and the mean local coordination of water molecules are slightly larger in the surface configuration of the cluster (Table 2), which is reflected in the difference between the $H_{rel}$ values (Table 3), which is nearly equal to an H-bond dissociation energy (ca. 7.5 kcal/mol) in favor of the surface structure. The mean Na-O distances for the five first-shell molecules are close in these two cluster configurations, namely 2.36 and 2.38 Å, which seems quite reasonable because in both variants it cannot be said that sodium is deeply immersed in a water structure.

At the larger number of water molecules, in $Na^+(H_2O)_{33}$ and $Na^+(H_2O)_{38}$ clusters, the surface and bulk configurations are already characterized by different *CN* values of sodium, by one larger in the bulk structures (Figs. 1, 5). In the transient $Na^+(H_2O)_{33}$ and both bulk and transient $Na^+(H_2O)_{38}$ cluster configurations, a half of the first-shell molecules are already completely immersed in the water structures, and only a half can be found close to the surface.

**Table 2.** Coordination number of sodium (*CN*), averaged Na-O distance for the *CN* molecules (<$r_{NaO}$>); number of hydrogen bonds ($N_{H-bonds}$); numbers of water molecules with different coordination neighborhoods (*N(dda)*, *N(ddaa)*, and *N(daa)*); mean number of bonds formed by one water molecule (<$N_{bonds}$>); and volumes (*V*) of Na$^+$(H$_2$O)$_n$ clusters with bulk, transient, and surface configurations

| n | kind | CN | <$r_{NaO}$>, Å | $N_{H-bonds}$ | N(dda) | N(ddaa) | N(daa) | <$N_{bonds}$> | V, Å$^3$ |
|---|---|---|---|---|---|---|---|---|---|
| 28 | surface | 5 | 2.359 | 45 | 7 | 9 | 9 | 3.4 | 486.43 |
|  | bulk | 5 | 2.380 | 44 | 6 | 9 | 8 | 3.3 | 489.62 |
| 38 | surface | 5 | 2.351 | 63 | 12 | 11 | 9 | 3.4 | 649.82 |
|  | transient | 6 | 2.464 | 62 | 11 | 13 | 10 | 3.4 | 651.47 |
|  | bulk | 5 | 2.464 | 63 | 14 | 9 | 12 | 3.5 | 651.13 |
| 43 | surface | 5 | 2.365 | 72 | 12 | 14 | 9 | 3.5 | 732.13 |
|  | transient | 5 (6) | 2.389 (2.474) | 71 | 10 | 17 | 6 | 3.4 | 735.39 |
|  | bulk | 6 | 2.462 | 73 | 11 | 17 | 9 | 3.5 | 730.10 |
| 46 | surface | 6 | 2.427 | 76 | 12 | 15 | 9 | 3.4 | 782.84 |
|  | transient | 6 | 2.448 | 79 | 13 | 18 | 11 | 3.6 | 779.67 |
|  | bulk | 6 | 2.448 | 78 | 14 | 16 | 12 | 3.5 | 779.19 |
| 51 | surface | 6 | 2.442 | 86 | 13 | 20 | 11 | 3.5 | 865.12 |
|  | transient | 5 (7) | 2.405 (2.637) | 89 | 16 | 19 | 13 | 3.6 | 865.06 |
|  | bulk | 6 | 2.471 | 88 | 14 | 20 | 11 | 3.6 | 861.55 |

Accordingly, the sodium coordination shell looks like a more or less distorted tetragonal bipyramid. When sodium is shifted toward the surface, its local neighborhood changes to the trigonal bipyramid, which may formally be treated as a lost of one first-shell molecule, and the whole cluster resembles a water bowl, in the concaved surface of which sodium is half countersunk, residing deeper in the larger cluster. In the latter case, only two water molecules of its first coordination shell are actually surface particles of the cluster with a *daa* kind of the neighborhood and an additional coordination bond to sodium. The differences in the total numbers of H-bonds are counterbalanced to a certain degree by the more complete coordination shells of sodium in the bulk configurations, which results in the very close adiabatic potential values of the configurations at the same molecular sizes of the clusters, much smaller compared to the H-bond dissociation energy. At the same time, the vibrational thermal energy increments in these systems are smaller in the case of the bulk structures, so that the resulting $H_{rel}/n$ values of the clusters are nearly the same (about 17.7 kcal/mol), with differences below the accuracy threshold of the method. Thus, if at n=28, the surface configuration seemed preferable both from the energetic and structural points of view, at n= 33 or 38, the surface (with *CN*=5), transient, and bulk (with *CN*=5 or 6) configurations are almost equiprobable, for the effects of opposite signs and nature are nearly cancelled out.

**Table 3.** The relative total electronic energies ($E_{rel}$, kcal/mol), vibrational contributions to the enthalpies under normal conditions ($H_{vib}$, kcal/mol), relative enthalpies obtained as their sums ($H_{rel}$, kcal/mol), and their normalized values ($H_{rel}/n$, kcal/mol) of Na$^+$(H$_2$O)$_n$ clusters with bulk, transient, and surface configurations

| n | kind | $E_{rel}$ | $H_{vib}$ | $H_{rel}$ | $H_{rel}/n$ |
|---|---|---|---|---|---|
| 28 | surface | 0 | 495.1 | 495.1 | 17.7 |
|  | bulk | 7.4 | 495.3 | 502.8 | 18.0 |
| 38 | surface | 0 | 672.7 | 672.7 | 17.7 |
|  | transient | 2.9 | 672.0 | 674.8 | 17.8 |
|  | bulk | 3.1 | 671.9 | 675.1 | 17.8 |
| 43 | surface | 13.7 | 761.4 | 775.0 | 18.0 |
|  | transient | 20.6 | 761.3 | 781.8 | 18.2 |
|  | bulk | 0 | 762.4 | 762.4 | 17.7 |
| 46 | surface | 27.1 | 813.3 | 840.4 | 18.3 |
|  | transient | 9.5 | 815.3 | 824.8 | 17.9 |
|  | bulk | 0 | 815.3 | 815.3 | 17.7 |
| 51 | surface | 20.3 | 902.9 | 923.2 | 18.1 |
|  | transient | 17.9 | 903.8 | 921.7 | 18.1 |
|  | bulk | 0 | 904.5 | 904.5 | 17.7 |

When the number of molecules is further increased to n=43, we see a different picture. Here, the first-shell of hexa-coordinate sodium in the bulk configuration is substantially asymmetric (with Na-O distances of 2.30, 2.32, 2.42, 2.43, 2.55, and 2.75 Å) and can be classified as intermediate between distorted tetragonal bipyramid and pentagonal pyramid. Even more asymmetric is the transient configuration of the cluster, in which five molecules are quite close to sodium, at a mean distance about 2.4 Å, while the sixth one is much more distant, at 2.9

Å, though judging from its orientation with respect to sodium and local coordination kind, it can (and probably should) be included in its first hydration shell. Depending on whether it is counted or not, we have here $CN$ equal to 5 or 6 and, accordingly, the mean Na-O distance is 2.389 or 2.474 Å respectively as shown in Table 2. The surface configuration involves a definitely penta-coordinate sodium and is characterized by the smallest mean $r_{NaO}$ distance. The energy difference between such configurations is quite large, which may be expected because of the different $CN$ values accompanied by the different numbers of hydrogen bonds (the largest in the bulk configuration and the smallest in the surface one). The difference between the bulk and surface configurations is nearly twice as large as the H-bond energy, while the energy of the transient structure is still higher by another H-bond energy increment. To our mind, this is a good illustration of the possible situation that may take place in solutions where the concentration of a sodium salt and the nature of the counterion predetermine the so-to-say transient hydration state of sodium between $CN = 5$ and 6. The change in the coordination type should be accompanied by quite a large change in the kinetic energy excess in the vibrational degrees of freedom. Moreover, the increase in the potential energy upon transformation of the bulk configuration into the transient-like one corresponds to the general compression of the whole structure: the apparent volume decreases from 744.3 to 735.4 Å$^3$, which is reasonable from the purely physical point of view.

Upon further increase in the number of water molecules to n= 46 and 51, the coordination number of sodium becomes steadily six in both surface and bulk configurations and varies around six in the transient ones, which at first sight seems a formally expected situation. Being located close to the center of a water nanodroplet in the bulk configurations, sodium finds itself coordinated to water molecules arranged in vertices of not as strongly distorted tetragonal bipyramid. When it is shifted toward the surface in the corresponding structures, its coordination neighborhood becomes more asymmetric. Typically one or two of the six Na-O coordination bonds are longer compared to the residual ones. In Na$^+$(H$_2$O)$_{46}$ and Na$^+$(H$_2$O)$_{51}$ clusters, the shorter and longer Na-O contacts are 2.34–2.45 vs. 2.49 Å and 2.34–2.47 vs. 2.49–2.57 Å respectively. And, here for the first time, one can see a larger (by two) number of hydrogen bonds between water molecules in the bulk configurations (Table 2), which means that the total number of molecules is already sufficient for the rational organization of the spatial structure, which, on one hand, is typical of water aggregates and, on the other, readily localizes sodium in its central part. The smaller the number of the molecules more distant from the sodium nucleus at the smaller corresponding distances and the larger the number of water–water hydrogen bonds, the larger the mean number of bonds per molecule (Table 2), which also reflects the more preferable spatial character of the bond networks. This conclusion is further supported by the smaller apparent volumes of the bulk structures, which means their relative compactness.

The structural preferences are reflected in the energetic characteristics of the cluster configurations. The bulk ones are definitely preferable judging from the relative adiabatic potential values, which are by 20 kcal/mol lower compared to those of the surface configurations (Table 3); and the difference cannot be leveled off by a slightly more preferable vibrational thermal contribution to the energy in the case of the surface structures, so that in the latter case the $H_{rel}/n$ values are about 18.1 to 18.3 kcal/mol compared to 17.7 kcal/mol typical of the bulk structures. The seemingly small difference of 0.4 to 0.6 kcal/mol is in fact large, if one takes into account that it should be multiplied by the number of molecules in the hydration structure.

In the corresponding quite a broad energy range between the surface and bulk configurations at such already large number of water molecules, many transient structures fall. Figure 6 shows two such configurations; and here their conditional name becomes completely feasible, because the coordination neighborhood of sodium is strongly asymmetric as if it appeared during some structure reorganization. These two transient configurations were selected for the relative closeness of the integral structural and energetic parameters of one of them to the surface configuration of the same cluster and of the other to the bulk one. Thus, all the range is formally spanned. In the transient $Na^+(H_2O)_{46}$ configuration, the shorter and longer Na-O distances are 2.32–2.46 and 2.50–2.64 Å respectively, while in $Na^+(H_2O)_{51}$, there are only five molecules at Na-O distances of 2.36–2.45 Å, while two surface molecules are as distant as at 3.11 and 3.32 Å. To a certain extent, the possibility of such sodium coordination is due to the more prolate cluster shape. It is interesting that not only the coordination number of sodium is the highest in the transient structures, but also the number of hydrogen bonds between water molecules is the largest among all the configurations considered. As a result, the mean coordination number of water molecules is also the largest, namely 3.6 (Table 2), which reflects the spatial reasonableness of such configurations. At the same time, the apparent volumes of the clusters with the transient configurations are closer to either the bulk or surface ones. And here, a clear correlation between this parameter and the relative stability of the configuration is observed, namely the smaller the difference between the volumes, the closer the energy values, and vice versa (Tables 2 and 3). However, it is worth noting that the total energy ranges, the bulk, transient, and surface configurations of the clusters fall in, are comparable to the relative enthalpies per molecule in all the cluster systems considered. And the above dynamic simulations showed that upon supplying such an excess kinetic energy to the cluster expectedly does not cause its even partial decomposition or strong reorganization, but rather promotes mutual transformations of the configurations due to the breathing of the whole structure, which can be absolutely complementary to similar changes in adjacent liquid-phase fragments but proceeding at a proper phase shift.

**CONCLUSIONS**

How can the results discussed above be summed up? First of all, there is a clear trend in the spatial organization of the hydration structure of sodium depending on the available amount of water molecules. When the mean number is smaller than 30, the coordination number of sodium equals five, and it prefers to reside close to a surface of the hydration structure. When the mean number of molecules increases above 30, but remains smaller than ca. 45, there appears a clear difference between the first-shell configuration of sodium, which is penta-coordinate in the surface structures, hexa-coordinate inside bulk structures, and has some intermediate coordination neighborhood in the transient configurations. Finally, when the number of molecules per ion increases above 45, at both bulk and surface variants of the sodium localization it is surrounded with six molecules, though their arrangement, as well as their effect as a hydration structure nucleus on the more distant molecules, is energetically more reasonable and preferable when sodium is completely immersed in the hydration structure rather than half countersunk in it. It is worth noting that in all the aforementioned configurations (either surface of bulk with either penta- or hexa-coordinate sodium), the $H_{rel}/n$ normalized relative thermal enthalpies are close on the average to 17.7 kcal/mol. When a comparable energy is accumulated within a structure fragment, which comprises about three dozens molecules around sodium, the internal reorganization of the hydration structure becomes possible due to the repeated changes in the local apparent volume and. hence, in the promoted oscillations of the first-shell molecules with respect to sodium.

When the corresponding mean hydration structures can be observed in real specimens? If we take into account that a part of ions are joined in contact or solvent separated pairs even at relatively low concentrations, we can assume that the actual number of molecules per ion can vary around the mean value, which equals 55.6/2C, where C is the molar concentration of the salt solution. In 1M solution, a mean number is close to 28, which corresponds to the smallest cluster considered. In such solutions (if we assume the nearly equal susceptibility of sodium and the counterion in the 1,1-electrolyte to attach and retain molecules in their solvation shells), the predominant coordination number of sodium should be five, and its position should be similar to those in the surface structures. At the higher concentrations, when the association of ions should be more pronounced, one can expect the coexistence of penta- and hexa-coordinate sodium. At the lower concentrations, up to 0.6 M, a similar oscillation between the possible penta-coordinate sodium in the surface position and hexa-coordinate sodium in the bulk may determine the actually recorded mean values. However, as we can see, here its is necessary to take into account the coexistence of these so-to-say ultimate situations along with any intermediate that can appear in the dynamically changing systems. Finally, when the concentration of an 1,1-electrolyte is lower than 0.5 M, one can believe that most of sodium ions are hexa-coordinate and quite well separated with the hydration shells from other ions.

**Table 4.** Mean distances ($<r_{NaO}>$, Å) in the first coordination shell of sodium in the clusters with bulk, transient, and surface configurations that involve penta- and hexa-coordinate sodium

| CN | cluster configuration | $<r_{NaO}>$ |
|---|---|---|
| 5 | surface and transient | 2.374 |
| 5 | bulk | 2.422 |
| 6 | surface and transient | 2.451 |
| 6 | bulk | 2.460 |

Judging from the results obtained, the mean Na-O distances should change consistently with the predominant coordination of sodium. If the structure information about the clusters considered is summed up, a clear correlation between the *CN* value, the cluster kind and the mean Na-O distance ($<r_{NaO}>$, Table 2) can be noticed. For example, at the same n, the $<r_{NaO}>$ value is smaller in the surface configuration, namely by 0.021 Å at *n*=28 and *CN*=5; 0.024 Å at *n*=43 and *CN*=5; 0.021 Å at *n*=46 and *CN*=6; and 0.029 Å at *n*=51 and *CN*=6. Concurrently, if one compares the $<r_{NaO}>$ values for the penta- and hexa-coordinate sodium at any numbers of water molecules in the cluster, again it becomes clear that the values are smaller at *CN*=5 and larger at *CN*=6. Mean Na-O distances at different coordination kinds in different hydration structures are summed in Table 4. As can be seen, $<r_{NaO}>$ = 2.374 and 2.422 Å at *CN*=5 in the surface and bulk variants of the cluster structures. Concurrently, the dynamic mean value obtained for $Na^+(H_2O)_{33}$ cluster is intermediate, namely 2.39 Å. Hence, probably distances of ca. 2.38–2.40 Å should be typical of relatively concentrated solutions. At the low concentrations, when sodium prefers to be hexa-coordinate, the mean Na-O distances increase to 2.451 and 2.460 Å at the surface and bulk localization, which may lead to an observable value around 2.45 Å.

What are the experimental estimates discussed in the Introduction? In fact, the recent XRD and EXAFS data were 2.38 and 2.37 Å respectively in 2.6 M solution [6], which probably reflects the preferable penta-coordinate state of sodium. At the same time, the dynamic Stokes radius of a sodium ion of 2.44 Å estimated in 0.5 to 6 M aqueous NaCl solutions with the pulsed field gradient NMR method [7] can be determined by the aforementioned coexistence of penta- and hexa-coordinate sodium at such concentrations and the discovered transient structures that can be considered as models of those intermediate configurations that can appear when penta- and hexa-coordinate shells transform into each other. The process can repeatedly take place on a timescale of 1 μs typical of the experiment, and the averaged value of those given in Table 2 is 2.42 Å. If one additionally takes into account that there may be repeated increases in the mean value, which are caused by the drift of some molecules to a larger distance from sodium (see Fig. 3, up to $<r_{NaO}>$ of 2.56 Å) or inclusion of additional (quite distant) molecules in the formal first solvation shell (as in transient configurations of $Na^+(H_2O)_n$ with n=43 and 51 where $<r_{NaO}>$ = 2.47 and 2.64 Å respectively), the resulting observed average value can be still higher than 4.42 Å.

It is also necessary to draw attention to the fact that different experimental techniques probe specimens to different depths; and, judging from the results discussed above, the higher the response fraction obtained from the surface layers, the larger the contribution of the surface-like hydration structures; and vice versa, the deeper the layers studied, the higher the contribution of the bulk-like structures to the experimental result. The contribution of the former kind increases also with a general increase in the concentration of the solution in question. Finally, the temporal resolution of the technique also has its implications on the final result for it always represents a certain averaging over the structures that may appear during the response generation and treatment. Thus, it cannot be unambiguously stated that some experimental estimates are more accurate than some others (typically previous). It is predetermined by the experimental techniques and setups and the concentrations of solutions studied; and the coordination numbers of sodium and the metal-oxygen distances can vary in relatively broad ranges, which reflect peculiarities of the inner hydration structures depending on the relative amount of water molecules available for their formation and the location of the structures, either deep in the bulk solution far from counterions or close to the counterions in the bulk part or to the surface.

**ACKNOWLEDGMENTS**

The research was supported by the Center for Laser Technologies and Materials Science, Institute of General Physics, Russian Academy of Sciences.

The data that support the findings of this study are available from the author upon reasonable request.